\documentclass[a4paper,12pt]{article}
\usepackage[colorlinks,citecolor=blue,linkcolor=blue,urlcolor=blue,filecolor=blue,backref=page]{hyperref}
\usepackage{amsfonts}
\usepackage{amsmath}
\usepackage{authblk}
\usepackage{ctable}
\usepackage[nottoc,numbib]{tocbibind}
\usepackage[round]{natbib}
\usepackage{enumitem}
\usepackage{float}
\usepackage[margin=1.05in]{geometry}
\usepackage{mathtools}
\usepackage{multirow}
\usepackage{nicefrac}
\usepackage{ragged2e}
\usepackage[labelfont=bf]{caption}
\usepackage{subcaption}
\usepackage[normalem]{ulem}
\usepackage{bbm}     
\usepackage{dsfont}  
\usepackage{graphicx}
\usepackage{pdflscape}

\title{\begin{center} Clustering high dimensional mixed data\\ to uncover sub-phenotypes:\\   joint analysis of phenotypic and genotypic data. \end{center}}

\begin{document}
\pagestyle{plain}
\date{}

\author[1]{Damien McParland}
\author[2]{Phillips, C.M.}
\author[3]{Brennan, L.}
\author[4]{Roche, H.M.}
\author[5]{Isobel Claire Gormley\thanks{claire.gormley@ucd.ie}}

\affil[1]{{\footnotesize School of Mathematics and Statistics, University College Dublin, Ireland.}}
\affil[2]{{\footnotesize HRB Centre for Diet and Health Research, Department of Epidemiology and Public Health, University College Cork, Ireland\\

HRB Centre for Diet and Health Research, School of Public Health, Physiotherapy and Sports Science, University College Dublin, Ireland\\

Nutrigenomics Research Group, UCD Conway Institute, University College Dublin, Ireland.}}
\affil[3]{{\footnotesize School of Agriculture and Food Science, UCD Institute of Food and Health, University College Dublin, Ireland.}}
\affil[4]{{\footnotesize Nutrigenomics Research Group, UCD Conway Institute, University College Dublin, Ireland.}}
\affil[5]{{\footnotesize  School of Mathematics and Statistics, University College Dublin, Ireland\\

INSIGHT: The National Centre for Data Analytics, University College Dublin, Ireland.}}

\maketitle

\newpage
\begin{abstract}
The LIPGENE-SU.VI.MAX study, like many others, recorded high dimensional continuous phenotypic data and categorical genotypic data. LIPGENE-SU.VI.MAX focuses on the need to account for both phenotypic and genetic factors when studying the metabolic syndrome (MetS), a complex disorder that can lead to higher risk of type 2 diabetes and cardiovascular disease. Interest lies in clustering the LIPGENE-SU.VI.MAX participants into homogeneous groups or \emph{sub-phenotypes}, by jointly considering their phenotypic and genotypic data, and in determining which variables are discriminatory. \\

 A novel latent variable model which elegantly accommodates high dimensional, mixed data is developed to cluster LIPGENE-SU.VI.MAX participants using a Bayesian finite mixture model.   A computationally efficient variable selection algorithm is incorporated, estimation is via a Gibbs sampling algorithm and an approximate BIC-MCMC criterion is developed to select the optimal model. \\
 
Two clusters or sub-phenotypes (`healthy' and `at risk') are uncovered. A small subset of variables is deemed discriminatory which notably includes phenotypic and genotypic variables, highlighting the need to jointly consider both factors. Further, seven years after the LIPGENE-SU.VI.MAX data were collected, participants underwent further analysis to diagnose presence or absence of the MetS. The two uncovered sub-phenotypes strongly correspond to the seven year follow up disease classification, highlighting the role of phenotypic and genotypic factors in the MetS, and emphasising the potential utility of the clustering approach in early screening. Additionally, the ability of the proposed approach to define the uncertainty in sub-phenotype membership at the participant level is synonymous with the concepts of precision medicine and nutrition.
\end{abstract}

\section*{Keywords}
clustering, mixed data,  phenotypic data, SNP data, metabolic syndrome.

 \section{Introduction}
 \label{sec:intro}
 
Many large cohort based studies collect high dimensional continuous phenotypic and categorical genotypic data. The pan European LIPGENE-SU.VI.MAX (SUpplementation en VItamines et Min\'{e}raux AntioXydants) study ({\tt www.ucd.ie/lipgene}) is one such study which focuses on the need to account for both phenotypic and genetic factors when studying the metabolic syndrome (MetS). The MetS is a complex disorder that can lead to increased risk of developing type 2 diabetes and cardiovascular disease. The MetS is the term used to describe a clustering of several risk factors for cardiovascular disease, namely obesity, abnormal blood lipids, insulin resistance and high blood pressure.  Obesity is on the rise globally and is considered to be a principle factor in the development of insulin resistance and the metabolic syndrome. The World Health Organisation estimates that the global prevalence of diabetes will almost double from 171 million people in 2000 to 300 million people by the year 2030. Given the strain this will place on health and health systems all over the world there is a need to gain greater understanding of the MetS, thereby reducing its adverse health effects. In particular, the influence of both phenotypic and genetic factors (and their interaction) on the MetS has recently come to the fore, and is the focus of the LIPGENE-SU.VI.MAX project. Valuable introductions and contributions to the LIPGENE-SU.VI.MAX project include \cite{phillips09a} and \cite{ferguson2010}.
 
 Under the LIPGENE-SU.VI.MAX study, high dimensional data of mixed type were collected on a group of participants. Continuous phenotypic variables (e.g. anthropometric and biochemical variables such as waist circumference and plasma fatty acid levels) as well as categorical (binary and nominal) genotypic single nucleotide-polymorphism (SNP) variables were recorded.  Here, interest lies in clustering the participants into homogeneous groups or \emph{sub-phenotypes}, based on jointly modelling their phenotypic and genotypic data, to uncover groups with similar phenotypic-genotypic profiles. In the LIPGENE-SU.VI.MAX study, a large number of phenotypic and genotypic variables were recorded; determining which variables discriminate between the resulting sub-phenotypes is therefore of interest. Moreover, given the ethos of the LIPGENE-SU.VI.MAX study, whether the set of discriminatory variables includes both phenotypic and genotypic variables is of key interest. The developed methodology has wide applicability beyond the LIPGENE-SU.VI.MAX study, in any setting seeking to uncover subgroups in a cohort on which high dimensional data of mixed type have been recorded. 
  
Joint modelling approaches for data of mixed type are gaining attention in a range of statistical and applied areas (see \cite{dunson05, faes08, wagner10, deLeon11, chen14}, among others, for example). In particular \cite{deLeon13} provides a comprehensive overview of recent methodological and applied advances in the mixed data modelling area. Latent factor models in particular have been successfully employed to jointly model mixed data;  \cite{quinn04,  gruhl13} and \cite{murray13} use factor analytic models to analyse mixed data but not in a clustering context. In a similar vein to the approach taken here, \cite{Huang2014} consider a joint analysis of SNP and gene expression data in studies of complex diseases such as asthma, but again not in the clustering context.  The MetS has had recent exposure in the statistical and computational literature -- \cite{Matsunaga2005, vattikuti2012} and \cite{Gostev2011} employ computational approaches to learn about the disease, but mainly from a genetic point of view. 
 
Latent variable based clustering models have been successfully utilised to analyse high dimensional data. For example, \cite{ghahramani97} propose a mixture of factor analysers model with a cluster specific parsimonious covariance matrix. A suite of similar models with varying levels of parsimony is developed in \cite{mcnicholas08} and the mixture of factor analysers model is fitted in a Bayesian framework in \cite{fokoue03}. Mixtures of structural equation models are developed in \cite{yung1997} and \cite{zhu2001}. More recent developments in this area include those in \cite{baek08}, \cite{baek10} and \cite{viroli2010}, among others. However, while these models can efficiently model high dimensional data, none of them can cluster observed mixed data while also correctly handling each variable type.
 
 Clustering data of mixed type is a challenging statistical problem. Early attempts to address the problem include the use of mixture models and location mixture models \cite{everitt88a,everitt88b, muthen99, hunt99, hunt03} as well as non-model based approaches \cite{huang97, ahmad07}; \cite{vermunt02} clusters mixed categorical data using a latent class analysis approach. More recently \cite{cai11, browne12, morlini11, cagnone12, gollini13} attempt to cluster mixed categorical data using latent variable models and \cite{biernacki15} cluster multivariate ordinal data using a stochastic binary search algorithm. However none of these can analyse the specific combination of continuous and categorical variables without transforming the original variables in some way, or can feasibly accommodate high dimensional data. An alternative model-based approach to clustering mixed continuous and categorical data, clustMD, is introduced in \cite{mcparland16}. While this approach can explicitly model the inherent nature of continuous and categorical variables directly, it is again computationally infeasible to use for high dimensional data. In particular, clustMD cannot accommodate large numbers of nominal variables. Copula models for clustering mixed data \cite{marbac14, kosmidis15} while showing distinct promise, also have limitations in high dimensional settings.

The recent mixture of factor analysers for mixed data (MFA-MD) \cite{mcparland14a} is a hybrid of latent variable models for different data types and provides the machinery for clustering mixed categorical data.  Here, the MFA-MD model is extended to facilitate clustering of high dimensional, mixed continuous and categorical data. Specifically, the joint model is composed of a factor analysis model for continuous data, an item response theory model for binary/ordinal data and a multinomial probit type model is used for nominal data. The clustering machinery is provided by a finite mixture model. 
 
 The MFA-MD model is ideal for high dimensional data settings as its factor analytic roots provide a parsimonious covariance structure. However large numbers of variables, as are present in the LIPGENE-SU.VI.MAX data, hamper the substantive interpretability of the resulting clusters and place a heavy computational burden on model fitting. Existing approaches to variable selection in a clustering context include reversible jump Markov chain Monte Carlo methods \cite{tadesse05}, approximate Bayes factors are used in \cite{raftery06} and \cite{maugis09} to compare nested sets of variables and \cite{wang08} use penalised model based clustering in the context of microarray data. Such methods would be computationally expensive given the latent variable aspect of the MFA-MD model, and given the large number of variables in LIPGENE-SU.VI.MAX data. Therefore, here an efficient novel online variable selection algorithm is incorporated when fitting the extended MFA-MD model, improving substantive interpretability and computational costs. Inspired by \cite{andrews14}, variable selection is based on a within cluster variance to overall variance criterion, efficiently leading to an interpretative clustering solution. Model fitting is performed in the Bayesian paradigm and is achieved via a Gibbs sampling algorithm. 
 
 As in any clustering setting, uncovering the number of underlying clusters is a key, and often difficult, question.  In the context of the extended MFA-MD model, the dimension of the latent factor aspect of the model also requires selection. Typical likelihood based model selection criteria such as the Bayesian Information Criterion (BIC) \cite{schwarz78, kass95} have been demonstrated to perform well in many general clustering settings \cite{fraley02, gormley06}, and marginal likelihood evaluation \cite{fruhwirth04} or the use of over fitting mixture models have gained warranted attention \cite{vanhavre15, malsinerwalli16} in the Bayesian literature. The likelihood function of the MFA-MD model is intractable however, rendering such approaches unusable. Therefore, here a novel approximation of the likelihood function is incorporated with the BIC-MCMC criterion \cite{fruhwirth11}, to efficiently select the optimal model (i.e. the optimal number of clusters and the optimal number of latent factors) in the context of the extended MFA-MD model.
 
 The extended MFA-MD model, with variable selection, is used to cluster the LIPGENE-SU.VI.MAX participants within a Bayesian framework. A range of models with varying numbers of clusters and latent factor dimensions are fitted. The BIC-MCMC criterion suggests two clusters or sub-phenotypes of participants, and a set of just 25 of the original 738 variables are deemed discriminatory. Examination of the cluster specific parameters reveals a `healthy' sub-phenotype and an `at risk' sub-phenotype. Notably the set of discriminatory variables contains both phenotypic and genotypic variables, highlighting the need to jointly consider both data types. Some of the discriminatory variables are intuitive and have been highlighted previously in the literature, but some of the discriminating SNPs in particular are novel discoveries.  
 
Seven years after the LIPGENE-SU.VI.MAX data analysed here were collected, each of the participants underwent further analysis to diagnose the presence or absence of the MetS, based on a criterion which considers continuous phenotypic data only. The two clusters uncovered here from the initial LIPGENE-SU.VI.MAX data strongly correspond to the seven year follow up disease classification, highlighting the role of phenotypic and genetic factors in the MetS and, perhaps most importantly, the potential utility of the clustering approach in early screening.
 
 The model-based nature of the MFA-MD approach to clustering provides a global view of the group structure in the set of LIPGENE-SU.VI.MAX participants. However, it additionally provides detailed insight to sub-phenotype membership at the participant level, through quantification of the probability of sub-phenotype membership for each participant. This ability to define the uncertainty of cluster membership is an important development for the application of the metabotyping concept in precision medicine and nutrition \cite{odonovan16}.

 The remainder of the paper is organised into the following sections. Section \ref{sec:data} provides background to the LIPGENE-SU.VI.MAX study, and specific details on the data collected. Section \ref{sec:methods} contains the three modelling contributions of the paper: (i) details of the extended MFA-MD model for high dimensional, mixed continuous and categorical data (ii) an outline of the variable selection and inference procedure and (iii) the development of the approximate BIC-MCMC model selection tool. Simulation studies, diverted to the Supplementary Material for clarity, provide evidence to support the modelling and selection approaches taken. Section \ref{sec:results} discusses the results of fitting the extended MFA-MD model for high dimensional data to the LIPGENE-SU.VI.MAX data, and considers model fit. The paper concludes in Section \ref{sec:discussion} with a discussion and some future research directions.
 
 \section{The LIPGENE-SU.VI.MAX study}
 \label{sec:data}
LIPGENE-SU.VI.MAX is a European Union Sixth Framework Integrated Programme entitled `Diet, genomics and the metabolic syndrome: an integrated nutrition, agro-food, social and economic analysis' conducted by 25 research centres across Europe. The primary focus of LIPGENE-SU.VI.MAX is the interaction of nutrients and genotype in the metabolic syndrome (MetS). The MetS is the term used to describe a clustering of several risk factors for cardiovascular disease, namely obesity, abnormal blood lipids (such as high blood cholesterol and raised triglyceride levels), insulin resistance and high blood pressure (hypertension). One quarter of the world's adult population have the metabolic syndrome and increasing numbers of children and adolescents have it as the worldwide obesity epidemic accelerates. Table \ref{tab:diagnosis} details the MetS diagnosis criterion used here which relates to insulin resistance, dyslipidaemia, cholesterol, blood pressure and abdominal obesity. Many closely related definitions of the MetS are also in use \cite{alberti05, alberti06, huang2009}.

  \begin{table}[h]
 \caption{ 
 A person with 3 or more of the abnormalities listed below is diagnosed as having the MetS.\label{tab:diagnosis}}
 \begin{center}
 \begin{tabular}{ll}\hline \hline
 Fasting glucose  &  $\geq$5.5 mmol l$^{-1}$ \\
 concentration & or treatment of previously diagnosed diabetes.\\\hline
 Serum TAG & $\geq$ 1.5 mmol l$^{-1}$ \\
 concentration & or treatment of previously diagnosed lipidemia.\\\hline
 Serum HDL-c &  $<$ 1.04 mmol l$^{-1}$ (Men) \\
 concentration & $<$ 1.29 mmol l$^{-1}$ (Women)\\\hline
 Blood pressure & Systolic BP $\geq 130$ mm Hg, Diastolic BP $\geq 85$ mm Hg\\
  & or treatment of previously diagnosed hypertension.\\\hline
 Waist & $> 94$ cm (Men), $>80$ cm (Women)\\
 Circumference & \\\hline
 \end{tabular}
 \end{center}
 \end{table}
 
Under LIPGENE-SU.VI.MAX, data from a prospective population-based study  were available \cite{ferguson2010, hercberg2004}. Twenty-six continuous phenotypic measurements in addition to 801 categorical SNP variables were recorded for each of 1754 participants. Examples of the continuous phenotypic measurements include fasting glucose concentration, waist circumference and plasma fatty acid levels. An example of a categorical genotypic variable is the nominal SNP {\tt rs512535} of the \emph{APOB} gene which has three genotypes, $AA$, $GG$ or $AG$ in the data. The 801 genotypic variables were selected using a candidate gene approach based on pathways adversely affected in the metabolic syndrome, and their relevant genes, as previously described in \cite{deEdelenyi08, phillips06}. Biological variables were based on characteristics of the metabolic syndrome \cite{alberti05, alberti06, huang2009} and plasma fatty acid profiles were determined as biomarkers of habitual dietary intake as previously described \cite{phillips09a}.

Some data cleaning was conducted prior to analysis.  Without loss of generality, the nominal SNP variables were coded with the convention 0 =  dominant homozygous, 1 = recessive homozygous and 2 = heterozygous. Any SNP variable with more than 100 missing values was removed, as were SNPs for which all 3 genotypes were not observed in the data (most of which only had one observed genotype and therefore are non-discriminatory in a clustering setting). A total of 990 participants were then removed as they still had at least one missing value across the remaining SNPs. 

Some of the remaining SNPs had a small number ($<$ 10\% of the number of participants) of counts of the recessive homozygous genotype. In such cases, for reasons of computational efficiency and stability, the recessive homozygous and the heterozygous categories were merged, thus resulting in some SNPs becoming binary variables. The merged category can be thought of as a `compound genotype'. For example, the SNP {\tt rs17777371} of the \emph{ADD1} gene became a binary SNP with genotypes $GG$ and $CG/CC$ in the data. While losing some information, merging of at least one sparsely observed genotype with another will not largely impact the findings in terms of uncovering clusters, or highlighting variables which discriminate between clusters. A total of 371 SNPs were collapsed to binary variables, leaving 341 nominal SNP variables. Finally the SNP data were combined with the continuous phenotypic data and participants that had any missing values for the continuous variables were removed. This left a final complete data set of 505 participants and 738 variables (26 continuous variables, 371 binary SNPs and 341 nominal SNPs); this data set is analysed here. No genotypes were removed solely as a result of missing data from other variables, and the continuous variables were standardised before any analysis was performed. The full list of 738 variables analysed here is given in the Supplementary Material.

 As stated LIPGENE-SU.VI.MAX was a prospective study. Seven years after the data analysed here were collected, new continuous phenotypic data were recorded on the LIPGENE-SU.VI.MAX participants in order to diagnose the presence or absence of the MetS, according to the criterion detailed in Table \ref{tab:diagnosis}. The correspondence between the clusters uncovered from the initial phenotypic and genotypic data and the seven year follow-up disease diagnosis is examined in Section \ref{sec:results}.

 \section{Modelling and inference}
 \label{sec:methods}
 
 A model-based approach is taken to cluster the LIPGENE-SU.VI.MAX participants, based on their initial mixed continuous, binary and nominal data. The mixture of factor analysers model for mixed ordinal and nominal data, MFA-MD, is introduced in \cite{mcparland14a}. Here the MFA-MD model is extended to also allow for continuous data, a variable selection procedure is proposed which facilitates feasible handling of high dimensional data, details of Bayesian inference are provided, and an approximate BIC-MCMC criterion for model selection is developed.

 \subsection{Modelling the continuous phenotypic variables}
 \label{subsec:cnsvars}
 
 A factor analysis model \cite{spearman1904} is used to model the multivariate continuous phenotypic data. Specifically, the observed $J$ continuous phenotypic measurements, $\underline{z}_i$, on participant $i$ are modelled as
 $$\underline{z}_i = \underline{\mu} + \Lambda \underline{\theta}_i + \underline{\epsilon}_i$$ 
 where $\underline{\mu}$ is a mean vector, $\Lambda$ is a loadings matrix and $\underline{\theta}_i$ is a participant specific latent trait. The error vector $\underline{\epsilon}_i$ follows a zero mean multivariate Gaussian distribution with diagonal covariance matrix $\Psi$. The dimension of the latent trait $\underline{\theta}_i$ is $Q$ where $Q \ll J$. The factor analysis model offers parsimony as the marginal covariance $\Sigma = \Lambda\Lambda^T + \Psi$ requires estimation of only $J(Q+1)$ parameters.

 \subsection{Modelling the binary SNP variables}
 \label{subsec:binSNPs}
 
As described in Section \ref{sec:data} some SNPs are treated as binary variables and are modelled using item response theory (IRT) models. Suppose that SNP {\tt rs17777371} is the $j^{th}$ variable (for $j = 1, \ldots, J$). IRT models assume that, for participant $i$, a latent Gaussian variable $z_{ij}$ corresponds to each observed binary response $y_{ij}$. A Gaussian link function is assumed, though other link functions, such as the logit, are detailed in the IRT literature \cite{lord68, fox10}. If $z_{ij} < 0$ then the binary response will be $y_{ij} = 0$ while if $z_{ij} > 0$ then $y_{ij} = 1$. Relating this to SNP {\tt rs17777371}, say, if $z_{ij} < 0$ then the observed genotype for participant $i$ will be $GG$ while if $z_{ij} > 0$ then observed genotype will be $CG/CC$.
 
 In a standard IRT model, a factor analytic structure is then used to model the underlying latent variable $z_{ij}$. It is assumed that the value of $z_{ij}$ depends on a $Q$ dimensional, participant specific, latent trait $\underline{\theta}_i$ (often termed the ability parameter) and on some variable specific parameters.  Specifically, the underlying latent variable $z_{ij}$ for respondent $i$ and variable $j$ is assumed to be distributed as 
 $$z_{ij}| \underline{\theta}_i \sim N(\mu_j + \underline{\lambda}_j^T\underline{\theta}_i, 1).$$ 
 The parameters $\underline{\lambda}_j$ and $\mu_j$ are usually termed the item discrimination parameters and the negative item difficulty parameter respectively. As in \cite{albert93}, a probit link function is used so the conditional variance of $z_{ij}$ is $1$.  Under this model, the conditional probability that $y_{ij}=1$ is 
 $$P(y_{ij}=1 |\mu_j,  \underline{\lambda}_j, \underline{\theta}_i) = \Phi\left(\mu_j + \underline{\lambda}_j^T\underline{\theta}_i\right)$$
 where $\Phi$ denotes the standard Gaussian cumulative distribution function.
 
 \subsection{Modelling nominal SNP variables}
 \label{nomSNPs}
 
 Modeling nominal data is challenging, due to the fact that the set of possible responses is not ordered. In the LIPGENE-SU.VI.MAX data, the possible responses for nominal SNPs is a set of three genotypes. For example, the nominal SNP {\tt rs512535} of the \emph{APOB} gene has three levels or genotypes, $AA$, $GG$ or $AG$, in the data. These response levels are coded as 0, 1 and 2 respectively, but no order is implied.
 
 As detailed in Section \ref{subsec:binSNPs}, the IRT model for binary SNP variables posits a one dimensional latent variable for each observed binary SNP. In the factor analytic model for nominal SNP variables, a two-dimensional latent \emph{vector} is required for each observed nominal SNP. That is, the latent vector for participant $i$ corresponding to nominal SNP $j$ is denoted $\underline{z}_{ij} = (z_{ij}^1, z_{ij}^{2})^T.$ The observed nominal response is then assumed to be a manifestation of the values of the elements of $\underline{z}_{ij}$ relative to each other and to a cut-off point, assumed to be $0$. That is,
  $$y_{ij} = \left\{ \begin{array}{ll}
          0 & \mbox{if $ \displaystyle \max\{z_{ij}^1, z_{ij}^{2}\} < 0$} \vspace{0.1cm}\\
          1 & \mbox{if $z_{ij}^{1} = \displaystyle \max\{z_{ij}^1, z_{ij}^{2}\}$ and $z_{ij}^{1} > 0$} \vspace{0.1cm}\\
          2 & \mbox{if $z_{ij}^{2} = \displaystyle \max\{z_{ij}^1, z_{ij}^{2}\}$ and $z_{ij}^{2} > 0$}.\\\end{array} \right. $$ 
 Similar to the IRT model, the latent vector $\underline{z}_{ij}$  is modelled via a factor analytic model. The mean of the conditional distribution of $\underline{z}_{ij}$ depends on a respondent specific, $Q$-dimensional, latent trait, $\underline{\theta}_i$, and item specific parameters i.e.
 $$\underline{z}_{ij}| \underline{\theta}_i \sim \mbox{MVN}_{2}(\underline{\mu}_j + \Lambda_j \underline{\theta}_i, \mathbf{I})$$
 where $\mathbf{I}$ denotes the identity matrix. The loadings matrix $\Lambda_{j}$ is a $2 \times Q$ matrix, analogous to the item discrimination parameter in the IRT model of Section \ref{subsec:binSNPs}; likewise, the mean $\underline{\mu}_{j}$ is analogous to the item difficulty parameter in the IRT model.
 
 It should be noted that binary data could also be regarded as nominal. The model proposed here is equivalent to the model proposed in Section \ref{subsec:binSNPs} when the number of possible levels is two.
 
 \subsection{A factor analysis model for mixed mixed continuous and categorical data}
 \label{subsec:famixed}
 
 The factor analysis model for continuous phenotypic variables, the IRT model for binary SNPs and the factor analysis model for nominal SNPs all have a common structure. These models are  combined to produce a unifying model for mixed continuous, binary and nominal data.
 
 For each participant $i$ there are $A = 26$ observed continuous phenotypic variables, $B = 371$ latent continuous variables corresponding to the binary SNP variables and $C = 341$ latent continuous vectors corresponding to the nominal SNPs. These are collected together in a single $D$ dimensional vector $\underline{z}_{i}$ where  $D = A + B + 2C$. That is, underlying participant $i$'s set of $J = 738 (=A+B+C)$ continuous, binary and nominal variables lies
 \begin{eqnarray*}
 \underline{z}_i  & = &\left(z_{i1}, \ldots, z_{iA}, z_{i(A+1)} \ldots, z_{i(A+B)}, z_{i(A+B+1)}^1, z_{i(A+B+1)}^2\ldots, z_{iJ}^1, z_{iJ}^{2} \right).
 \end{eqnarray*}
 The first $A$ entries of this vector are the observed continuous measurements. The remaining entries are latent data underlying the categorical responses. This vector is then modelled using a factor analytic structure i.e. 
 $$\underline{z}_{i}| \underline{\theta}_i   \sim  \mbox{MVN}_{D}(\underline{\mu} + \Lambda\underline{\theta}_i,  \Psi).$$ The $D \times Q$ dimensional matrix $\Lambda$ is termed the loadings matrix and $\underline{\mu}$ is the mean vector. The entries of the diagonal covariance matrix $\Psi$ are 1 along the diagonal, with the exception of the first $A$ entries which correspond to the continuous variables.
 
 This model provides a parsimonious factor analysis model for the high dimensional latent vector $\underline{z}_{i}$ which underlies the observed mixed data. Marginally the latent vector is distributed as
 $\underline{z}_{i} \sim \mbox{MVN}_{D}(\underline{\mu}, \Lambda \Lambda^{T} + \Psi)$
 resulting in a  parsimonious covariance structure for $\underline{z}_{i}$.

 \subsection{A mixture of factor analyzers model for mixed continuous and categorical data}
 \label{subsec:HybMix}
 
 To facilitate clustering, the hybrid model defined in Section \ref{subsec:famixed} is placed within a mixture modeling framework resulting in the extended mixture of factor analyzers model for mixed data (MFA-MD). In the MFA-MD model, clustering occurs at the latent variable level. That is, under the MFA-MD model the distribution of the observed and latent data $\underline{z}_i$ is modeled as a mixture of $G$ Gaussian densities i.e. 
 $$f(\underline{z}_i)  =  \sum_{g=1}^{G}\pi_g \mbox{MVN}_{D}\left( \underline{\mu}_g,  \:\:  \Lambda_g \Lambda_g^T +\Psi \right).$$ 
 The probability of belonging to cluster $g$ is denoted by $\pi_g$ ($\sum_{g=1}^{G}\pi_g = 1$, $\pi_g > 0$ $\forall$ $g$). The mean and loadings are cluster specific, while $\Psi$ is equal across clusters for additional parsimony. Constraining the loadings matrices to be equal across clusters, similar in ethos to the mixture of common factor analysers \cite{baek08, baek10}, would offer further parsimony but result in a subtly yet importantly different model. 
 
 As is standard in a model-based approach to clustering \cite{mclachlan88, fraley98, mclachlan00, celeux00, fraley02}, a latent indicator variable, $\underline{\ell}_i = (\ell_{i1}, \ldots, \ell_{iG})$ is introduced for each participant $i$. This binary vector indicates the cluster to which participant $i$ belongs i.e. $l_{ig} = 1$ if $i$ belongs to cluster $g$; all other entries in the vector are $0$. 
 
 Under the MFA-MD model for mixed continuous and categorical data, the augmented likelihood function for the $N = 505$ participants is
 \begin{eqnarray}\nonumber
 & &\prod_{i=1}^{N} \prod_{g=1}^{G} \left\{ \pi_g \left[ \prod_{j=1}^{A} N(z_{ij}
 |\tilde{\underline{\lambda}}_{gj}^{T} \tilde{\underline{\theta}}_{i}, \psi_{jj})\right] \right.\\ \nonumber
 && \left. \times \left[ \prod_{j=A+1}^{B} \prod_{k=0}^{1 }N^{T}(z_{ij}
 |\tilde{\underline{\lambda}}_{gj}^{T} \tilde{\underline{\theta}}_{i}, 1)^{\mathbb{I}
 \{y_{ij} = k\}} \right] \right . \\
 && \left. \times \left[\prod_{j=A+B+1}^{J} \prod_{k=1}^{2} \prod_{s=0}^{2} N^T(z_{ij}^{k} |
 \tilde{\underline{\lambda}}_{gj}^{k^{T}} \tilde{\underline{\theta}}_{i}, 1)^{\mathbb{I}(y_{ij}=s)}\right] \right\}^{\ell_{ig}}\label{eqn:like}
  \end{eqnarray}
  where $\tilde{\underline{\theta}}_i = (1, \theta_{i1}, \ldots, \theta_{iq})^T$ and $\tilde{\Lambda}_g$ is the matrix resulting from the combination of
 $\underline{\mu}_g$ and $\Lambda_g$ so that the first column of $\tilde{\Lambda}_g$ is $\underline{\mu}_g$. In the binary part of the model, the Gaussian is truncated between $-\infty$ and 0 if $y_{ij} = 0$, and between 0 and $\infty$ otherwise. In the nominal part of the model, The Gaussian is also truncated, depending on the observed $y_{ij}$ i.e. 
 \begin{itemize}
  \item If $y_{ij}=0$ then $ \displaystyle \max\{z_{ij}^1, z_{ij}^{2}\} < 0$.
  \item If $y_{ij}=1$ then $z_{ij}^{1} = \displaystyle \max\{z_{ij}^1, z_{ij}^{2}\}$ and $z_{ij}^{1} > 0$, $z_{ij}^2$ is restricted so that $z_{ij}^2 < z_{ij}^{1}$.
  \item If $y_{ij}=2$ then $z_{ij}^{2} = \displaystyle \max\{z_{ij}^1, z_{ij}^{2}\}$ and $z_{ij}^{2} > 0$, $z_{ij}^1$ is restricted so that $z_{ij}^1 < z_{ij}^{2}$.
 \end{itemize}

 The MFA-MD model proposed here is related to the mixture of factor analyzers model \cite{ghahramani97, mclachlan00} which is appropriate when the observed data are all continuous in nature. A Bayesian treatment of such a model is detailed in \cite{fokoue03}; \cite{mcnicholas08} detail a suite of parsimonious mixture of factor analyzer models.
 
 \subsection{Variable selection}
 \label{subsec:varsel}
 
 The LIPGENE-SU.VI.MAX data contain a large number of variables, particularly categorical variables. A variable selection algorithm that removes variables which have no clustering information would ease the computational burden of the model fitting process and also provide substantive interpretation advantages by only retaining variables which discriminate between clusters.
 
 A simple but effective online variable selection procedure is incorporated here. For an informative or discriminatory variable, the within cluster variance will be lower than the overall variance for that variable in the data. Variables for which the within cluster and overall variances are similar do not discriminate between clusters and are not interesting in a clustering context. Specifically, for each variable $j$, a variance ratio $VR_{j}$ is computed where
 \begin{eqnarray}
 {VR}_j & = & \displaystyle \sum_{g=1}^{G}\sum_{\substack{i=1\\ \forall i \in g}}^{n_g}(z_{ij} - \bar{z}_{gj})^2 / \sum_{i=1}^N(z_{ij} - \bar{z}_j)^2.
 \label{eqn:vrj}
 \end{eqnarray}
 The variance ratio is computed in an online manner in that at an iteration of the model fitting algorithm $n_g$ denotes the number of participants currently classified as members of cluster $g$. In turn, the empirical cluster mean for cluster $g$ and variable $j$ is denoted by $\bar{z}_{gj}$, while the overall mean for variable $j$ is denoted by $\bar{z}_j$. 
 
 Small values for $VR_j$ indicate that variable $j$ discriminates between clusters while larger values indicate that variable $j$ takes similar values across all clusters and therefore contains no clustering information. A user specified threshold $\varepsilon$ is set such that if $VR_j > \varepsilon$, variable $j$ is dropped from the model and otherwise it is retained. Selection of $\varepsilon$ is application specific and its choice within the LIPGENE-SU.VI.MAX analysis is discussed in Section \ref{sec:results}. The choice of $\epsilon$ can be thought of as the choice of how many variables the model will highlight as discriminatory; $\epsilon$ doesn't have an `optimal' value as is typical of many tuning parameters. Decreasing $\epsilon$ is equivalent to indicating that a more aggressive variable selection is desirable. This variable selection method is shown to perform well in simulation studies, provided in the Supplementary Material.
 
 \subsection{Bayesian inference}
 \label{subsec:ParEst}
 
 The Bayesian paradigm is a natural framework for the estimation of latent variable models. Fitting the proposed MFA-MD model in a Bayesian framework requires specification of prior distributions for all parameters. Conjugate prior distributions are employed here. Specifically,  $ \tilde{\underline{\lambda}}_{gd} \sim  \mbox{MVN}_{(Q+1)}(\underline{\mu}_{\lambda}, \Sigma_{\lambda}), \hspace{0.2cm} \underline{\pi} \sim \mbox{Dir}(\underline{\alpha})$ and $\psi_{jj} \sim \mathcal{G}^{-1}(\beta_1, \beta_2)$. For participant $i$, it is assumed the latent trait $\underline{\theta}_i$ follows a standard multivariate Gaussian distribution while the latent indicator variable $\underline{\ell}_i$ follows a Multinomial$(1, \underline{\pi})$ distribution. Further, conditional on membership of cluster $g$, the latent variable $\underline{z}_{i}| l_{ig} = 1 \sim \mbox{MVN}_{D}(\underline{\mu}_{g}, \Lambda_{g} \Lambda_{g}^{T} + \mathbf{\Psi})$. Combining these latent variable distributions and prior distributions with the augmented likelihood function specified in (\ref{eqn:like}) results in the joint posterior distribution, from which samples of the model parameters and latent variables are drawn using a Gibbs sampling MCMC scheme.
 
 Full conditional distributions for the latent variables and model parameters are detailed below; derivations and definitions of the distributional parameters are given in the Supplementary Material.
 
 \begin{itemize}
  \item Allocation vectors: $\underline{\ell}_i|\ldots \sim \mbox{Multinomial}(1,\underline{p})$.
 \item Mixing proportions:  $\underline{\pi} | \ldots \sim \mbox{Dirichlet}(\underline{\delta}_\pi)$.
 \item Latent traits:  $\underline{\theta}_i| \ldots \sim \mbox{MVN}_Q\left( \underline{\mu}_\theta, \Sigma_\theta \right)$.
 \item Item parameters: $\tilde{\underline{\lambda}}_{gd} | \ldots \sim \mbox{MVN}_{(q+1)} \left(\underline{\zeta}_\lambda, \Omega_\lambda\right)$.
 \item Error variance parameters: $\psi_{jj} \sim \mathcal{G}^{-1}(b_{1j}, b_{2j})$.
 \end{itemize}
 The full conditional distribution for the latent data $\mathbf{z}$ follows a truncated Gaussian distribution. The point of truncation depends on the form of the corresponding variable, the observed response, and the previously sampled values of $\mathbf{z}$ in the MCMC chain. The distributions are truncated to satisfy the conditions detailed in Section \ref{subsec:HybMix}. The latent variable $z_{ij}$ is therefore updated as detailed below. Note that $z_{ij}$ is not sampled for $j = 1, \ldots, A$ as in the case of the continuous variables $y_{ij} = z_{ij}$.
 \begin{itemize}
 \item If variable $j$ is binary and $y_{ij}=0$ then $z_{ij}| \ldots \sim N^T\left(\tilde{\underline{\lambda}}_{gj}^T\tilde{\underline{\theta}}_i, 1 \right)$
 where the distribution is truncated on the interval $(-\infty, 0)$. The truncation interval is $(0, \infty)$ if $y_{ij}=1$.
 
 \item If item $j$ is nominal then $z_{ij}^k | \ldots \sim N^T \left( \tilde{\underline{\lambda}}_{gj}^{k^T}\tilde{\underline{\theta}}_i, 1\right)$
 where $\tilde{\underline{\lambda}}_{gj}^k$ is the row of $\tilde{\Lambda}_g$ corresponding to $z_{ij}^k$ and the truncation intervals are defined as follows:
 \begin{itemize}
 \item if $y_{ij}=0$ then $z_{ij}^k \in (-\infty,0)$  for $k = 1,2$.
 \item if $y_{ij}=k$ for $k=1,2$ then:
 \begin{enumerate}
 \item $z_{ij}^{k} \in (\tau, \infty)$ where $\tau = \max\left(0, \displaystyle \max_{l \neq k}\{z_{ij}^l\}\right)$.
 \item for $l \neq k$ then $z_{ij}^{l} \in \left( - \infty, z_{ij}^{k}\right)$.
 \end{enumerate}
 \end{itemize}
 \end{itemize}
 Note that in the case of $y_{ij} = k \neq 0$,  the value $z_{ij}^{l}$ in the evaluation of $\tau$ in step 1 is the previously sampled value in the MCMC chain. The value of $z_{ij}^{k}$ in step 2 is the value sampled in step 1.
 
 The variable selection method presented in Section \ref{subsec:varsel} is incorporated into the outlined Gibbs sampler, and thus the proposed MFA-MD model is fitted in three stages:
 \begin{enumerate}
  \item \emph{Burn in phase}: In the first phase of the model fitting procedure all variables are included and the Gibbs sampling algorithm is run until convergence. 
  
  \item \emph{Variable selection phase}: After the burn in phase the algorithm moves into the variable selection phase. Given the current clustering, the variance ratio $VR_j$ is computed for each variable $j$. All variables for which $VR_j$ is greater than $\varepsilon$ are dropped from the model. The algorithm is allowed to burn in again before another variable selection step is performed, with a user specified frequency. The variable selection phase ends when no variables are removed from the model at a number of successive variable selection steps.
  
  \item \emph{Posterior sampling phase}: During this phase the Gibbs sampling algorithm proceeds, given the discriminating variables.
 \end{enumerate}
 
 Given its factor analytic roots, the MFA-MD model is not identifiable. Here, the loadings matrices are unconstrained and a Procrustean rotation is employed to solve the problem of their rotational invariance, following ideas in \cite{hoff02}, \cite{handcock07} and as detailed in \cite{mcparland14a}. Further, the well known clustering label switching problem is addressed using a loss function approach as in \cite{stephens00}.
 
 \subsection{Model selection via an approximate BIC-MCMC criterion}
 \label{sec:modelselection}
 
 As with any clustering problem, the number $G$ of clusters is unknown. Moreover, in the case of the MFA-MD model, the dimension of the latent trait $Q$ is also unknown. Under a model based approach to clustering, such as that taken here, the use of principled, statistical model selection tools to choose both $G$ and $Q$ are available. 
 
 Formal likelihood based criteria such as the Bayesian Information Criterion (BIC) \cite{schwarz78, kass95} have been demonstrated to perform well in many general clustering settings (e.g. \cite{fraley02, gormley06}), and also in clustering settings involving latent factor models (e.g. \cite{mcnicholas08}) and variable selection (e.g. \cite{raftery06, houseman2008}). There is also a rich Bayesian literature regarding model evidence; model selection tools based on the marginal likelihood \cite{friel12, fruhwirth06, mcparland13}
 are a natural approach to general model selection within the Bayesian paradigm, with reversible jump MCMC methods \cite{richardson97} and Markov birth-death methods \cite{stephens00b} popular in the context of clustering. More recently, overfitting approaches to model selection within clustering using Bayesian finite mixtures have gained warranted attention \cite{vanhavre15, malsinerwalli16}.  In the context of choosing $Q$ in latent factor models, \cite{lopes04} provide a comprehensive overview of Bayesian model assessment.  
 
Such approaches naturally require evaluation of the joint likelihood of the observed continuous and categorical data $Y$ which for the MFA-MD model is intractable as it requires integrating a multidimensional truncated Gaussian distribution, where truncation limits differ and are dependent across the dimensions. These approaches also require the variables in the data to be the same when comparing models. Thus, in order to select the optimal MFA-MD model an approximation of the observed data likelihood is constructed which involves both variables retained and removed during the variable selection steps. 
 
 Recall that for participant $i$ their observed data consists of $A$ continuous phenotypic variables, $B$ binary SNP variables and $C$ nominal SNP variables collected in $\underline{y}_i = (y_{i1}, \ldots, y_{iJ})$ where $J = A + B + C.$ Denoting the $\ddot{A}$ continuous, $\ddot{B}$ binary and $\ddot{C}$ nominal variables with clustering information collectively as $\underline{\ddot{y}}_{i}$ and the $\dot{A}$ continuous, $\dot{B}$ binary and $\dot{C}$ nominal variables with no clustering information collectively by $\underline{\dot{y}}_{i}$, the contribution to the likelihood function for participant $i$ is approximated as
 \begin{eqnarray}\nonumber
 \mathcal{\tilde{L}}_i & = & f(\underline{\ddot{y}}_{i})  f(\underline{\dot{y}}_{i})\\\nonumber
  & = & \left[ \sum_{g=1}^{G} \pi_g \left\{ \mbox{MVN}_{\ddot{A}}(\mu_g, \Lambda_{g} \Lambda_{g}^{T} + \Psi) \prod_{j=1}^{\ddot{B}+\ddot{C}} P(\ddot{y}_{ij}| i \in g)  \right\}\right] \\
   &  & \times \left[ \mbox{MVN}_{\dot{A}}(\mu, \Lambda \Lambda^{T} + \Psi) \prod_{j=1}^{\dot{B}+\dot{C}} P(\dot{y}_{ij}) \right].
 \label{eqn:approxl}
 \end{eqnarray}
 That is, independence is first assumed between the set of discriminating and the set of non-clustering variables. Further, for the discriminating variables, conditional independence between the set of $\ddot{A}$ continuous and the set of $\ddot{B}+\ddot{C}$ categorical variables is assumed, and within the set of $\ddot{B}+\ddot{C}$ categorical variables. Additionally, independence between the set of continuous and the set of categorical variables without clustering information is also assumed, and within the set of non-clustering categorical variables. 
 
 The multivariate Gaussian densities for the continuous variables in (\ref{eqn:approxl}) are straight forward to evaluate; a single Bayesian factor analysis model is fitted to the $\dot{A}$ removed variables. For the categorical variables in (\ref{eqn:approxl}), simple empirical probabilities are calculated from the observed data. For the $\ddot{B}+\ddot{C}$ categorical discriminating variables, these probabilities are the observed response probabilities, within each cluster. For the $\dot{B}+\dot{C}$ non-clustering categorical variables, the probabilities are the observed response probabilities among the $N$ participants. Thus a tractable approximation to the intractable likelihood is available, and can be used to compare models with varying values of $G$ and $Q$, and with varying sets of discriminating variables.
 
 This approximated observed likelihood function is incorporated in the BIC-MCMC \cite{fruhwirth11} criterion to perform model selection with the MFA-MD model. Analogous to the traditional BIC, the BIC-MCMC is derived from the largest observed log likelihood value generated across the MCMC draws, penalised for lack of parsimony. In the context of the MFA-MD model the approximate BIC-MCMC is defined as:
 
 $$\mbox{BIC-MCMC } = 2 \times \log \mathcal{\tilde{L}} - \nu \times \log(N)$$
 where  $\mathcal{\tilde{L}}  =  \prod_{i=1}^{N} \mathcal{\tilde{L}}_{i}$ denotes the largest observed approximate likelihood value across the MCMC draws and $\nu$ denotes the number of parameters in (\ref{eqn:approxl}). Thus for $G = 1, \ldots, G_{\max}$ and $Q = 1, \ldots, Q_{\max}$, the approximate observed likelihood function $\mathcal{\tilde{L}}$ is evaluated at each MCMC iteration, and the largest value used to compute the associated BIC-MCMC. The model with largest BIC-MCMC is chosen as the optimal model. The BIC-MCMC has been shown to perform well in the context of mixture models generally \cite{fruhwirth11}; its performance in combination with the likelihood approximation within the MFA-MD context is also shown to perform well in the simulation studies provided in the Supplementary Material.

 \section{Results}
 \label{sec:results}
 
\begin{figure}[ht]
 \centerline{\includegraphics[height=0.5\textheight,width=0.9\textwidth]{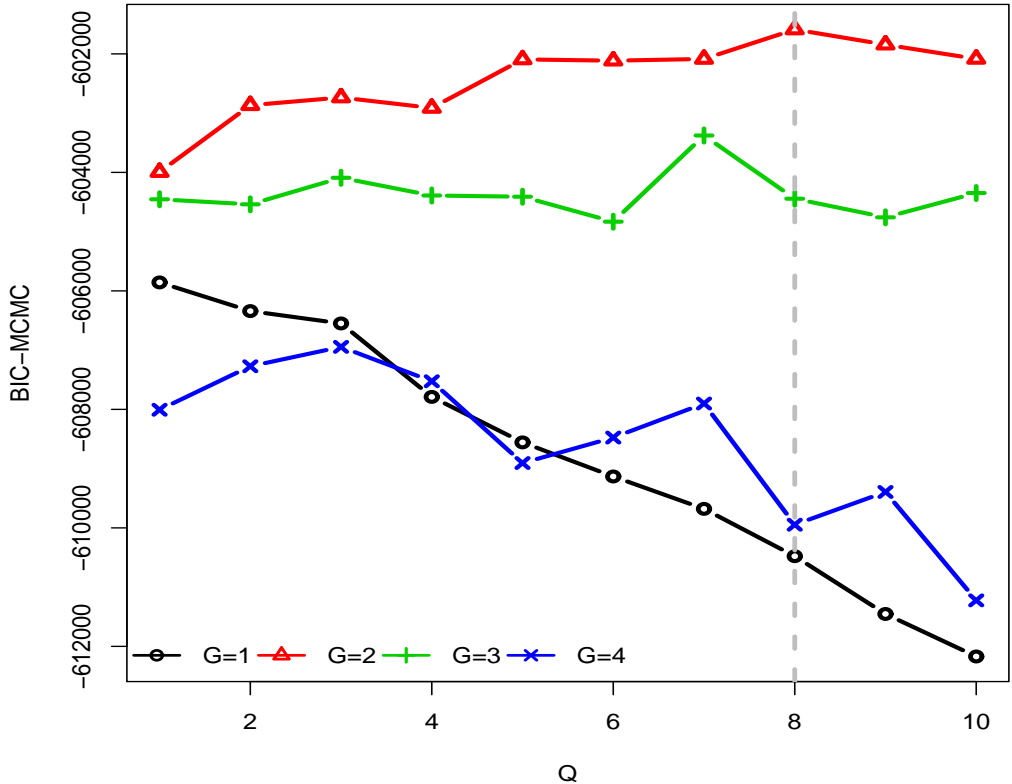}}
 \caption{The approximate BIC-MCMC for each of the MFA-MD models fitted to the set of LIPGENE-SU.VI.MAX participants. The dashed grey line indicates the largest approximate BIC-MCMC value achieved; the optimal model has two clusters and eight latent dimensions. }\label{fig:BICMCMC}
 \end{figure}

In order to cluster the set of LIPGENE-SU.VI.MAX participants, a number of MFA-MD models with $G=1, \ldots, G_{\max} = 4$ and $Q=1, \ldots, Q_{\max} = 10$ were fitted to the initial mixed phenotypic and genotypic data. The maximum value considered for $G$ was motivated by expert opinion on the expected structure of the set of participants; the maximum value of $Q$ considered was motivated by the observed performance of the $G = 1$ model (see Figure \ref{fig:BICMCMC}), and by run time considerations. The Jeffreys prior, Dirichlet$(0.5, \ldots, 0.5)$, was specified for the mixing proportions $\underline{\pi}$. An inverse gamma prior, with shape and scale parameters of 7, was specified for the $A$ diagonal elements of $\Psi$ corresponding to continuous variables. The mode of this relatively uninformative prior is just less than 1. A zero mean multivariate Gaussian prior was specified for $\tilde{\underline{\lambda}}_{gd}$ with $\Sigma_\lambda = 5\mathbf{I}$, which again is relatively uninformative. Prior sensitivity was assessed by trialling different values of the hyperparameters. The results were relatively insensitive to changes in the hyperparameters for $\underline{\pi}$ and $\tilde{\underline{\lambda}}_{gd}$ but somewhat sensitive to the hyper parameters for $\psi_{jj}$. Sensitivity to these inverse gamma hyperparameter values is a known problem for Bayesian inference of models of this type \cite{gelman06}. 
 
 For each of the forty models fitted, the \emph{burn in phase} was run for 20,000 iterations and in the \emph{variable selection phase} the variance ratio criterion was computed every 1000 iterations. This period between variable selection steps allowed the MCMC algorithm to `burn in' again after variables have been removed. In the LIPGENE-SU.VI.MAX setting, the variable selection threshold $\varepsilon$ was fixed at $0.95$ for continuous phenotypic variables and at $0.99$ for categorical SNP variables. These thresholds are very conservative so that only the most uninformative variables were removed. The thresholds could be lowered to facilitate a more aggressive variable selection procedure. The model fitting algorithm remained in the \emph{variable selection phase} until no variables were removed from the model for four successive variable selection steps. When this occurred the algorithm moved into the \emph{posterior sampling phase} which was then run for 100,000 iterations, thinned every $100^{th}$ iteration. Convergence of the Markov chains was assessed using trace and auto-correlation plots. Computation times for these models are variable  as the speed will depend on how many variables are removed and on both the dimension of the latent trait and the number of clusters fitted to the data. The $G=1$, $Q=1$ model took approximately 11 hours while the $G=4$, $Q=10$ model took approximately 25 hours. It should be noted that no variable selection can be applied if only one cluster is fitted to the data. The optimal model described below took less than 5 hours to fit as only a small number of variables were deemed discriminatory. These timings were measured by fitting the model using one processor of a quad core (2.83GHz) desktop PC with 4GB of RAM.
 
 The optimal MFA-MD model was selected using the approximate BIC-MCMC criterion developed in Section \ref{sec:modelselection}. Figure \ref{fig:BICMCMC} illustrates the approximate BIC-MCMC for each of the forty models fitted; the optimal model is indicated to have  $G = 2$ clusters and $Q = 8$ latent factors.  During fitting of the optimal $G = 2, Q=8$ MFA-MD model, a large number of variables is dropped at the beginning of the \emph{variable selection phase} but as the phase proceeds the model converges on a relatively small number of discriminatory variables. A plot showing the evolution of the number of variables retained during the \emph{variable selection phase} is given in the Supplementary Material. Only 25 of the original 738 variables are retained under the $G = 2, Q = 8$ model. Of those retained, 12 are continuous phenotypic variables, 2 are binary SNP variables and 11 are nominal SNP variables. Notably, in the nearest competing model $G = 2, Q = 9$, a total of 22 variables were retained, 16 of which were the same as those retained in the optimal $G = 2, Q = 8$ model; this pattern was observed in general within models with the same number of groups. 

Alternative variable selection criteria to (\ref{eqn:vrj}) are possible: the set of 40 models were also fitted using a weighted version of $VR_j$ where each squared difference in the numerator in (\ref{eqn:vrj}) is multiplied by the posterior probability that observation $i$ belongs to cluster $g$ and a `fuzzy' clustering version of the cluster specific means $\bar{z}_{gj}$ is also used. This fuzzy version of $VR_j$ allows observations that are not assigned to a cluster with a high degree of certainty to contribute to the within cluster variances of multiple clusters. In the case of the LIPGENE-SU.VI.MAX study, it was found that this fuzzy version of $VR_j$ had no effect on the optimal models: the same variables were chosen and the same clustering solutions were found, thus giving the same interpretation.
  
Of particular note, in the context of the LIPGENE-SU.VI.MAX study, is that both phenotypic and genotypic variables are deemed to be informative. The reduction from 738 to 25 variables aids the substantive interpretation of the model output significantly and ensures model fitting efficiency. Examination of the cluster specific parameters under the optimal model provides insight to the clustering structure in the set of LIPGENE-SU.VI.MAX participants; posterior inferences from the optimal MFA-MD model are discussed in what follows.

\subsection{Examining the cluster specific parameters for the set of discriminatory variables}
 
The reduced cardinality of the set of variables facilitates interpretation of the substantive differences between the resulting clusters or `sub-phenotypes'. 
 
 \begin{figure}[htp]
 \begin{center}
 \includegraphics[height = 0.5\textheight, width=0.9\textwidth]{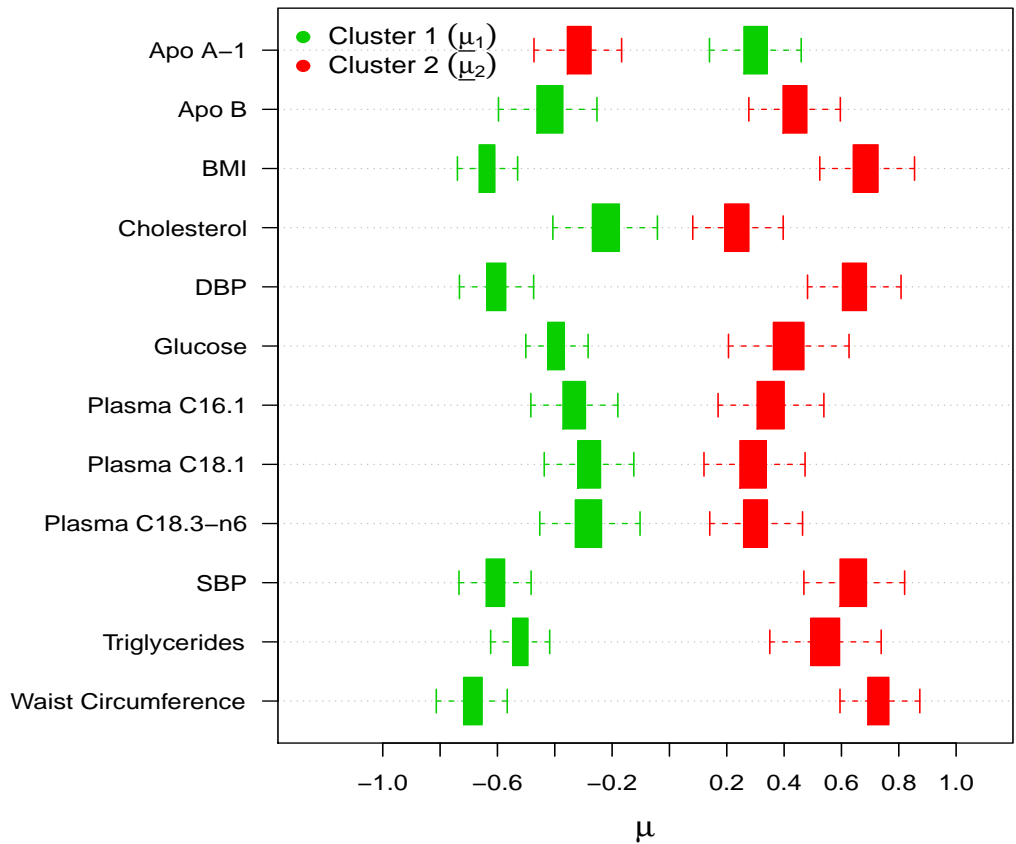}
 \caption{Box plots of the MCMC samples of mean parameters in each cluster, for the discriminating continuous phenotypic variables. All variables were standardised prior to analysis. The original units for each variable are detailed in the Supplementary Material.}
 \label{fig:cnsBox}
 \end{center}
 \end{figure}

%
 
 \begin{table}
\caption{Empirical posterior probabilities of each retained SNP genotype conditional on cluster membership. Associated uncertainties are all less than 0.01. 
}
\label{tab:PostProb}
\centering
\begin{subtable}{.5\textwidth}
\caption{\emph{ADD1} ({\tt rs17777371})}
\centering
\begin{tabular}{c|ccc}
  \hline
  & $GG$ & $CG/CC$ &  \\ \hline
  Cluster 1 & 0.80 & 0.20 &  \\ 
  Cluster 2 & 0.91 & 0.09 &  \\ \hline
\end{tabular}
\end{subtable}
\begin{subtable}{.5\textwidth}
\caption{\emph{APOB} ({\tt rs512535})}
\centering
\begin{tabular}{c|ccc}
  \hline
  & $GG$ & $AA$ & $AG$ \\ \hline
Cluster 1 & 0.22 & 0.23 & 0.55 \\ 
  Cluster 2 & 0.30 & 0.23 & 0.47 \\ \hline
\end{tabular}
\end{subtable}
\\

\begin{subtable}{.5\textwidth}
\caption{\emph{APOL1} ({\tt rs136147})}
\centering
\begin{tabular}{c|ccc}
  \hline
  & $CC$ & $AA$ & $AC$ \\ \hline
Cluster 1 & 0.33 & 0.21 & 0.46 \\ 
  Cluster 2 & 0.21 & 0.32 & 0.48 \\ \hline
\end{tabular}
\end{subtable}
\begin{subtable}{.5\textwidth}
\caption{\emph{CETP} ({\tt rs4784744})}
\centering
\begin{tabular}{c|ccc}
  \hline
  & $GG$ & $AA$ & $AG$ \\ \hline
Cluster 1 & 0.52 & 0.10 & 0.38 \\ 
  Cluster 2 & 0.32 & 0.11 & 0.57 \\ \hline
\end{tabular}
\end{subtable}
\\

\begin{subtable}{.5\textwidth}
\caption{\emph{FABP1} ({\tt rs2970901})}
\centering
\begin{tabular}{c|ccc}
  \hline
  & $CC$ & $AA$ & $AC$ \\ \hline
  Cluster 1 & 0.25 & 0.18 & 0.57 \\ 
  Cluster 2 & 0.37 & 0.17 & 0.46 \\ \hline
\end{tabular}
\end{subtable}
\begin{subtable}{.5\textwidth}
\caption{\emph{GYS1} ({\tt rs2270938})}
\centering
\begin{tabular}{c|ccc}
  \hline
  & $TT$ & $AA$ & $AT$ \\ \hline
  Cluster 1 & 0.34 & 0.21 & 0.45 \\ 
  Cluster 2 & 0.35 & 0.13 & 0.52 \\ \hline
\end{tabular}
\end{subtable}
\\

\begin{subtable}{.5\textwidth}
\caption{\emph{INSIG1} ({\tt rs9770068})}
\centering
\begin{tabular}{c|ccc}
  \hline
  & $CC$ & $TT$ & $TC$ \\ \hline
  Cluster 1 & 0.32 & 0.13 & 0.55 \\ 
  Cluster 2 & 0.35 & 0.20 & 0.45 \\ \hline
\end{tabular}
\end{subtable}
\begin{subtable}{.5\textwidth}
\caption{\emph{LRP2} ({\tt rs2544377})}
\centering
\begin{tabular}{c|ccc}
  \hline
  & $GG$ & $AA$ & $AG$ \\ \hline
  Cluster 1 & 0.48 & 0.11 & 0.41 \\ 
  Cluster 2 & 0.44 & 0.08 & 0.48 \\ \hline
\end{tabular}
\end{subtable}
\\

\begin{subtable}{.5\textwidth}
\caption{\emph{OLR1} ({\tt rs1050289})}
\centering
\begin{tabular}{c|ccc}
  \hline
  & $GG$ & $AG/AA$ &  \\ \hline
  Cluster 1 & 0.75 & 0.25 &  \\ 
  Cluster 2 & 0.83 & 0.17 &  \\ \hline
\end{tabular}
\end{subtable}
\begin{subtable}{.5\textwidth}
\caption{\emph{SLC25A14} ({\tt rs2235800})}
\centering
\begin{tabular}{c|ccc}
  \hline
  & $TT$ & $AA$ & $AT$ \\ \hline
  Cluster 1 & 0.38 & 0.35 & 0.26 \\ 
  Cluster 2 & 0.60 & 0.30 & 0.11 \\ \hline
\end{tabular}
\end{subtable}
\\

\begin{subtable}{.5\textwidth}
\caption{\emph{SLC27A6} ({\tt rs185411})}
\centering
\begin{tabular}{c|ccc}
  \hline
  & $GG$ & $AA$ & $AG$ \\ \hline
  Cluster 1 & 0.46 & 0.07 & 0.47 \\ 
  Cluster 2 & 0.47 & 0.16 & 0.37 \\ 
\end{tabular}
\end{subtable}
\begin{subtable}{.5\textwidth}
\caption{\emph{SLC6A14} ({\tt rs2071877})}
\centering
\begin{tabular}{c|ccc}
  \hline
  & $GG$ & $AA$ & $AG$ \\ \hline
  Cluster 1 & 0.63 & 0.17 & 0.19 \\ 
  Cluster 2 & 0.70 & 0.23 & 0.08 \\ \hline
\end{tabular}
\end{subtable}
\\

\begin{subtable}{.5\textwidth}
\caption{\emph{THYN1} ({\tt rs570113})}
\centering
\begin{tabular}{c|ccc}
  \hline
  & $GG$ & $AA$ & $AG$ \\ \hline
  Cluster 1 & 0.40 & 0.12 & 0.48 \\ 
  Cluster 2 & 0.48 & 0.09 & 0.43 \\ \hline
\end{tabular}
\end{subtable}
\end{table}

 The means of the retained continuous phenotypic variables for each cluster are illustrated in Figure~\ref{fig:cnsBox}. Examination of these posterior parameter estimates provides particular insight to the structure of the two clusters. Cluster 1 appears to be a `healthy' sub-phenotype in that the phenotypic variable means are lower in general in cluster 1 than in cluster 2.  It is well known that lower values of such phenotypic variables are typically associated with better health. For example, the mean levels of triglycerides, waist circumference, body mass index (BMI) and systolic and diastolic blood pressure variables (SBP and DBP respectively) are notably lower in cluster 1 than cluster 2. The exception is Apo A-1, the major structural protein of the high density lipoprotein (HDL) particle, low levels of which are a recognised risk factor for cardiovascular disease \cite{wilson98, gordon89}.  Apo A-1 levels are usually low when HDL cholesterol levels are reduced, thus it is intuitive that higher Apo A-1 levels are reported in the healthy cluster.   
 
Table~\ref{tab:PostProb} details the empirical posterior probability of each genotype across the thirteen retained SNPs, conditional on cluster membership. Clear differences in the distributions between clusters are visible. For example, in both retained binary SNPs {\tt rs17777371} of the \emph{ADD1} gene and {\tt rs1050289} of the \emph{OLR1} gene participants in both clusters are most likely to take the dominant homozygous genotype. However, for both SNPs, cluster 1 is more likely to take the compound recessive homozygous/heterozygous genotype (the second level) than cluster 2. In terms of retained nominal SNPs, the probability distributions between clusters for the {\tt rs4784744} SNP of the \emph{CETP} gene and the {\tt rs2235800} SNP of the \emph{SLC25A14} gene also show some disparities, for example. For the {\tt rs4784744} SNP of the \emph{CETP} gene, participants in cluster 1 are more likely to have the dominant homozygous genotype than those in cluster 2, with those in cluster 2 more likely to have the heterozygous genotype. For the {\tt rs2235800} SNP of the \emph{SLC25A14} gene, 60\% of participants assigned to cluster 2 have the dominant homozygous genotype compared to 38\% of those in cluster 1. The probability distribution is much more evenly spread across the genotypes for participants in cluster 1 than for those in cluster 2.

The 13 SNP variables deemed to be discriminatory are also listed in Table~\ref{tab:SNPdetails}, which provides details on characteristics of the discriminating SNPs and the biological pathways to which they are associated. Most of the SNPs deemed to be discriminatory are involved in lipid metabolism, glucose homeostasis or blood pressure regulation.  Associations between polymorphisms of a number of genes involved in fatty acid and lipid metabolism, inflammation, appetite control and adiposity with risk of the MetS or its features have previously been identified in the LIPGENE-SU.VI.MAX cohort \cite{phillips12b, phillips12a, phillips11a, phillips10c, phillips10b, phillips10a, phillips10d,  phillips09a, phillips09b, phillips09c}; some of these SNPs are also highlighted here, in addition to some novel discoveries.  

Of particular interest in the current analysis is the \emph{APOB} {\tt rs512535} SNP which has previously been reported to have association with MetS risk \cite{phillips11a}. Apo B is the main apolipoprotein associated with low density lipoprotein and the triglyceride rich lipoproteins \cite{chan92}. Other findings of note are {\tt rs9770068} of the \emph{INSIG1} gene which is involved in cholesterol metabolism \cite{radhakrishnan07} 
and {\tt rs4784744} of the \emph{CETP} gene which is involved in mediating exchange of lipids between lipoproteins and reverse cholesterol transport \cite{kuivenhoven98}; {\tt rs2544377} of the \emph{LRP2} gene and the {\tt rs1050289} SNP of the \emph{OLR1} gene, both of which are involved in lipid homeostasis \cite{lillis08,  sawamura97}; {\tt rs2970901} of the \emph{FABP1} gene and {\tt rs185411} of the \emph{SLC27A6} gene both of which are involved in fatty acid metabolism \cite{pelsers03, auinger12} and {\tt rs17777371} of the \emph{ADD1} gene which is involved in blood pressure regulation \cite{barlassina97}. 
 
%
 
Examination of the posterior parameter estimates across all discriminating variables suggests that cluster 1 could be termed a `healthy' sub-phenotype and cluster 2 an `at risk' sub-phenotype. Further, some of the phenotypic and SNP variables deemed to be discriminatory appear intuitive, while others are suggestive of potentially interesting relationships for further research.

\begin{landscape}
 \begin{table}[h]
 \begin{center}
 \caption{Characteristics of the set of 13 binary and nominal SNP variables deemed to be discriminatory.(Source: NCBI SNP data base \url{http://www.ncbi.nlm.nih.gov/SNP/})}
 \label{tab:SNPdetails}
 \begin{tabular}{lllll}\hline \hline
 Gene	& SNP & SNP type & Chromosome & Associated biological\\
 &&&& pathway\\\hline
  \emph{ADD1}  &{\tt rs17777371}& Adducin 1 & Flanking\_3UTR	& Blood pressure \\
  & && 	 chromosome 4 & regulation \\\hline
 \emph{APOB}  & {\tt rs512535} &	Apolipoprotein B& 	Intronic chromosome 2	& Lipid metabolism\\ \hline
   \emph{APOL1} & {\tt rs136147} & Apolipoprotein L1	 & Intronic chromosome 22 &	Lipid metabolism\\ \hline
 \emph{CETP}  & {\tt rs4784744} & Cholesterol ester transfer protein &  Intronic chromosome 16& Lipid metabolism\\ \hline
 \emph{FABP1}  & {\tt rs2970901} & 	Fatty acid binding protein 1, & Flanking\_5UTR & Lipid metabolism\\ 
 & & liver & chromosome 2 & \\\hline
   \emph{GYS1}  & {\tt rs2270938} &Glycogen synthase 1& 	Intronic chromosome 19	&  Glucose homeostasis\\ \hline
  \emph{INSIG1} & {\tt rs9770068} & Insulin Induced Gene 1 &  Intronic chromosome 7& Lipid metabolism, \\
 &&&&innate immunity. \\ \hline
 \emph{LRP2}  & {\tt rs2544377} & 	LDL receptor related protein 2	& Intronic chromosome 2	& Lipid metabolism\\ \hline
 \emph{OLR1} &  {\tt rs1050289} & Oxidized low density & 3UTR chromosome 12 &Lipid metabolism \\
 &&  lipoprotein (lectin-like) & &\\
  &&  receptor 1	& 	&\\\hline  
  \emph{SLC25A14}  & {\tt rs2235800} & Solute Carrier Family 25 &Intronic  x chromosome & Oxidative \\
 & &  (Mitochondrial Carrier, Brain), &  & phosphorylation\\
  && Member 14 or UCP5 & & \\ \hline
  \emph{SLC27A6} & {\tt rs185411} &Solute Carrier Family 27 	&Intronic chromosome 5 & Lipid metabolism\\
 &  & (Fatty acid transported), member 6 & & \\ \hline
   \emph{SLC6A14} & {\tt rs2071877} &	Solute carrier family 6 & Intronic  x chromosome & Amino acid \\
  &&  (amino acid transporter), & & transporter \\
  && member 14 &  		& \\\hline
 \emph{THYN1}	& {\tt rs570113} & Thymocyte nuclear protein 1 &Intronic chromosome 11 &Amino acid \\
 & & & & metabolism\\ \hline
 \end{tabular}
 \end{center}
 \end{table}
 \end{landscape}

\subsection{Correspondence between sub-phenotype membership and the seven year follow-up MetS diagnosis}
 
 As stated, the data analysed here are an initial set of measurements under the LIPGENE-SU.VI.MAX study. At a seven year follow-up, new continuous phenotypic data on each of the 505 participants were recorded. Each participant was then diagnosed as having the MetS or not based on the criterion in Table~\ref{tab:diagnosis}, which considers continuous phenotypic data only. It is therefore of interest to compare the cluster or sub-phenotype membership of each LIPGENE-SU.VI.MAX participant based on their initial phenotypic and genotypic data to their subsequent MetS diagnosis, seven years later.
 
The cluster or sub-phenotype membership for each participant is obtained by first computing the conditional probability that participant $i$ belongs to each cluster based on the MCMC samples, and a `hard' clustering is then obtained by assigning each participant to the cluster for which they have largest membership probability. Table~\ref{tab:rand} details the cross tabulation of the initial sub-phenotypes and the follow-up MetS diagnosis. It can be seen that traits of the MetS are apparent in the initial data, as the cross-tabulation shows good agreement, with a Rand index of 0.73 (and an adjusted Rand index of 0.46). Notably, Figure \ref{fig:BICMCMC} suggests there are five closely competing models to the optimal $G = 2, Q = 8$ model i.e. the $G = 2, Q = 5, 6, 7, 9, 10$ models. Comparing the resulting clusterings from these models to the follow-up MetS diagnosis results in Rand indices ranging from 0.71 to 0.74 and in adjusted Rand indices ranging from 0.42 to 0.48, suggesting that the models deemed optimal by the BIC-MCMC criterion all indeed have similar performance and perform well.
 
 \begin{table}[ht]
 \caption{Cross tabulation of sub-phenotype membership (based on fitting the MFA-MD model to the initial phenotypic and genotypic data) and MetS diagnosis (based on the diagnosis criterion in Table \ref{tab:diagnosis} on seven year follow up phenotypic data only). The Rand index is 0.73 (adjusted Rand index = 0.46).}
 \label{tab:rand}
 \begin{center}
 \begin{tabular}{llll}
  &&\multicolumn{2}{c}{Follow up data} \\
  && Healthy & MetS  \\ \hline
 \multirow{ 2}{*}{Initial data}  & Cluster 1 (`Healthy') & 220 &  42 \\ 
  & Cluster 2 (`At risk') & 39 & 204 \\\hline
 \end{tabular}
 \end{center}
 \end{table}
 
 Of further interest is whether the level of correspondence between the sub-phenotypes and the follow-up MetS diagnosis is stronger than that observed between the MetS diagnoses from both time points based on the phenotypic data only. One of the abnormalities required for diagnosis involves HDL cholesterol -- HDL cholesterol data are not available in the initial measurements however. Therefore the current diagnosis criterion in Table~\ref{tab:diagnosis} cannot be applied to the initial data. Hence, participants are diagnosed as MetS cases if they satisfy two or more of the remaining four diagnostic conditions relating to waist circumference, blood pressure, TAG and glucose concentration.  Table ~\ref{tab:MetS_t0} details the cross tabulation of the `initial diagnosis' compared to the `follow-up diagnosis' based on the phenotypic data only. Notably, the follow up diagnosis does not change here if it is based on 2 of the 4 available variables rather than on the criterion outlined in Table~\ref{tab:diagnosis}. Table ~\ref{tab:MetS_t0} also suggests that the traits of the MetS are apparent in the initial data, as the MetS diagnoses from the two time points agree well, with a Rand index of 0.69 (adjusted Rand index of 0.38). However, the level of agreement is lower in Table ~\ref{tab:MetS_t0} than that observed in Table~\ref{tab:rand}, highlighting the importance of utilising \emph{both} phenotypic and genotypic factors, and the potential utility of the clustering approach in early screening.

 \begin{table}[ht]
 \caption{Cross tabulation of MetS diagnoses from initial and follow up data. The Rand index is 0.69 (adjusted Rand index is 0.38)}
 \label{tab:MetS_t0}
 \begin{center}
 \begin{tabular}{llll}
  &&\multicolumn{2}{c}{Follow up data} \\
  && Healthy & MetS  \\ \hline
 \multirow{ 2}{*}{Initial data}  & Healthy & 194 &  31 \\ 
  & MetS & 65 & 215 \\\hline
 \end{tabular}
 \end{center}
 \end{table}
 
Further, to explore the influence of modelling each data type in its innate form, a $k$-means clustering algorithm with $k = 2$ was applied to all the 738 variables, treating all the SNP variable codes as continuous values. Comparing the resulting clustering to the follow up MetS diagnosis gave a Rand index of 0.60 (adjusted Rand index = 0.21). Applying $k$-means clustering (again with $k = 2$) to the set of 25 variables selected as discriminatory under the optimal $G = 2, Q = 8$ model gave a Rand index of 0.68 (adjusted Rand index = 0.37) when compared to the follow-up MetS diagnosis. As noted the MFA-MD model achieved a Rand index of 0.73 (adjusted Rand index = 0.46) highlighting the benefit of modelling the variables in their innate form.

Finally, the MFA-MD model outlined above was fitted to only the continuous phenotypic variables from the initial LIPGENE-SU.VI.MAX data. The optimal model, according to the approximate BIC-MCMC, was the $G = 2, Q = 7$ model, which gave a Rand index of 0.50 (adjusted Rand of 0.005) with the follow-up MetS diagnosis. This model under-performs when compared to analysing the phenotypic and genetic data jointly, again highlighting the importance of considering phenotypic and genotypic factors simultaneously with regard to early screening for the MetS.
 
\subsection{Quantifying uncertainty in sub-phenotype membership at the participant level}
 
One of the main advantages of a model-based approach to clustering is the inherent assessment of the uncertainty about cluster membership \cite{bensmail97, gormley06}. In the LIPGENE-SU.VI.MAX context, the model-based approach allows quantification of the probability of sub-phenotype membership for each participant. As stated, the cluster membership for each participant is obtained by first computing the conditional probability that participant $i$ belongs to each cluster based on the MCMC samples, and a `hard' clustering is then obtained by assigning each participant to the cluster for which they have largest membership probability. The uncertainty with which participant $i$ is assigned to its cluster may then be estimated by
\[U_i = \min_{g=1,\ldots, G} \{1 - \mathbf{P}(\mbox{cluster } g \ | \ \mbox{participant }i)\}. \]
If participant $i$ is strongly associated with cluster $g$ then $U_i$ will be close to zero.  

\begin{figure}[htp]
    \centering
    \begin{subfigure}{0.5\textwidth}
    \centering
        \includegraphics[width=0.95\linewidth]{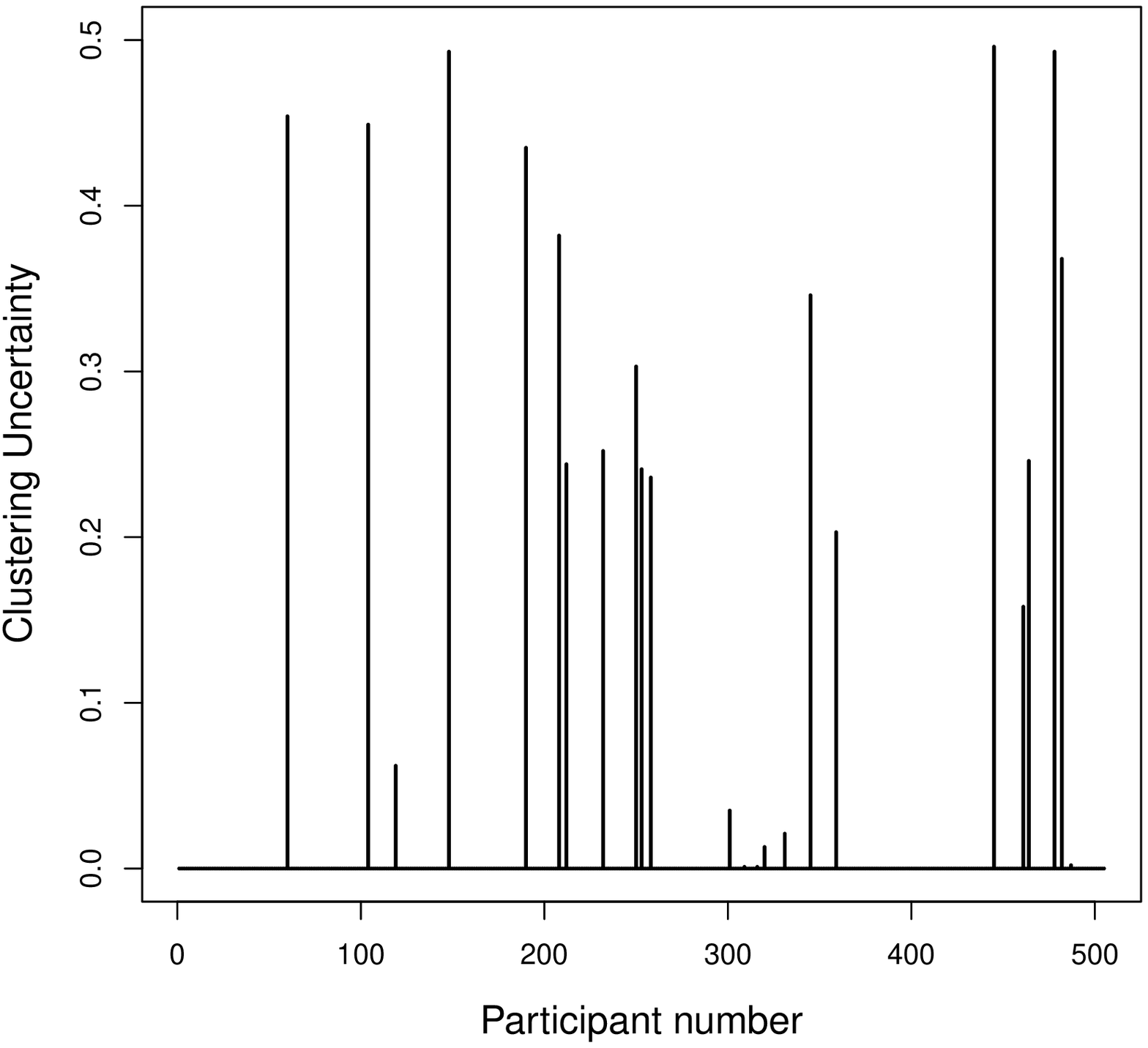}
        \caption{}
        \label{fig:UncertB}
    \end{subfigure}%
    \begin{subfigure}{0.5\textwidth}
    \centering
        \includegraphics[width=0.95\linewidth]{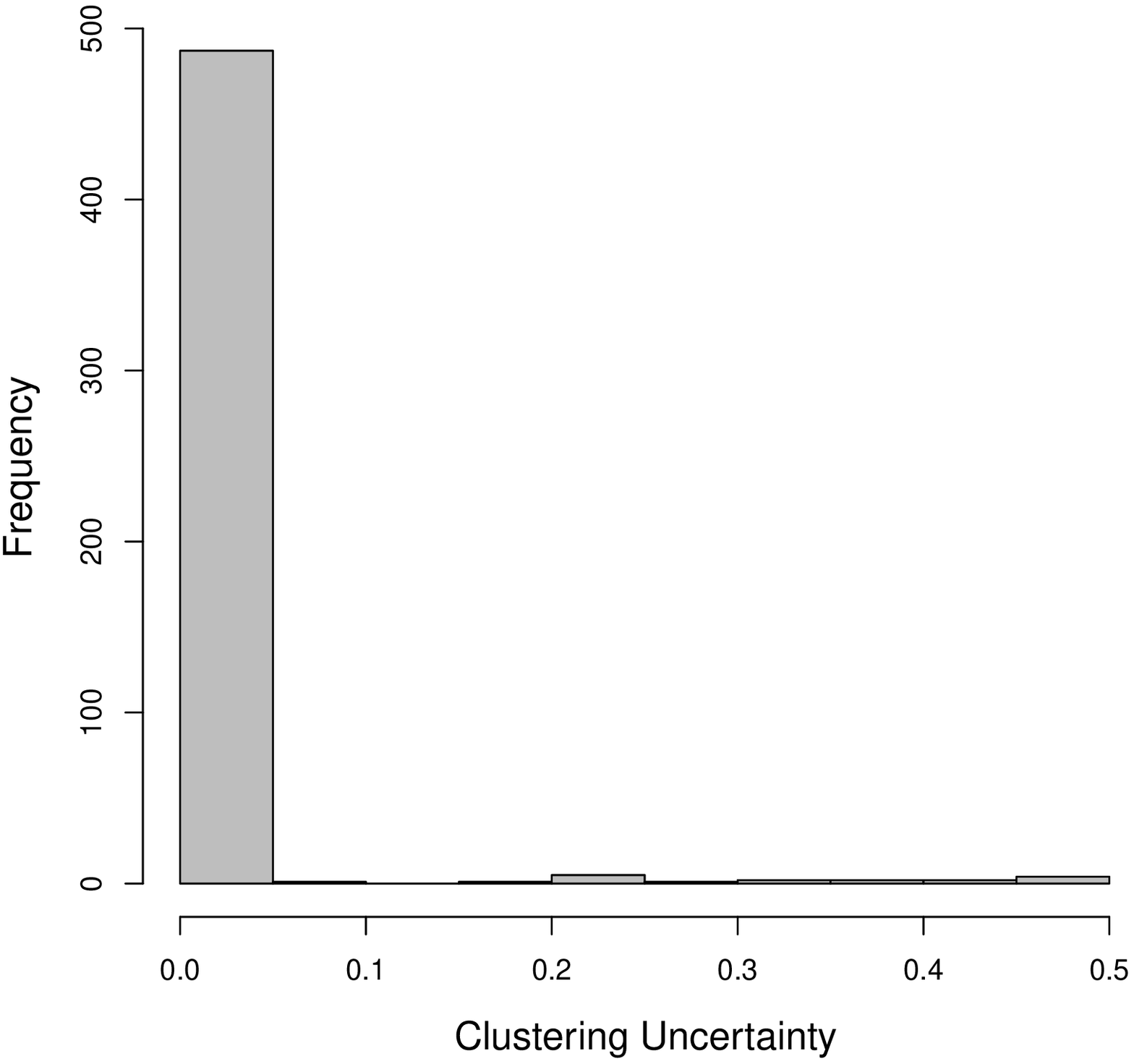}
        \caption{}
        \label{fig:UncertA}
    \end{subfigure}
    \caption{(a) The participant specific clustering uncertainties and (b) the histogram of the clustering uncertainties across all participants, under the optimal MFA-MD model.}
    \label{fig:Uncert}
    \end{figure}

Figure \ref{fig:Uncert} illustrates the clustering uncertainties under the optimal MFA-MD model. Figure \ref{fig:UncertB} illustrates the clustering uncertainty for each LIPGENE-SU.VI.MAX participant. The maximum uncertainty observed is 0.496, associated with participant  number 445. This participant is clustered with the `healthy' sub-phenotype, but there is high uncertainty associated with this clustering. Examination of this participant's data provides insight to this high clustering uncertainty -- participant 445 has much higher SBP and DBP, and much lower Apo A-1 levels than the mean levels in the `healthy' sub-phenotype. Further, participant 445 differs from the modal genotypes observed in the `healthy' sub-phenotype for SNPs \emph{APOB} ({\tt rs512535}), \emph{FABP1} ({\tt rs2970901}) and \emph{INSIG1} ({\tt rs9770068}). Thus while this participant is clustered with the `healthy' sub-phenotype they have large probability of being `at risk'. 

Thus, the model-based nature of the MFA-MD approach to clustering provides a global view of the group structure in the LIPGENE-SU.VI.MAX participants, but also provides detailed insight to sub-phenotype membership at the participant level; the ability to define the uncertainty in cluster membership is an important development for the application of the metabotyping concept in precision medicine and nutrition \cite{odonovan16}. Overall, the vast majority of LIPGENE-SU.VI.MAX participants have very small clustering uncertainty, as illustrated by Figure \ref{fig:UncertA}.


 \subsection{Assessing model fit}
 
In order to assess how well the selected MFA-MD model fits the LIPGENE-SU.VI.MAX data, Bayesian residuals and Bayesian latent residuals are utilised \cite{johnson99, fox10}. 

 For continuous phenotypic variables the Bayesian residual for participant $i$ on variable $j$ is 
 $$\epsilon_{ij} = \left(z_{ij} - \tilde{\underline{\lambda}}_{gj}^T\tilde{\theta}_i\right)/\psi_{jj}.$$
 The continuous phenotypic data are observed so this residual may be calculated explicitly by subtracting $\tilde{\underline{\lambda}}_{gj}^T\tilde{\theta}_i$ at each MCMC iteration from $z_{ij}$ and dividing this quantity by $\psi_{jj}$ from that iteration. For a well fitting model, this residual follows a standard Gaussian distribution.
 
 However, $z_{ij}$ corresponding to a categorical SNP variable is not observed but sampled during the MCMC scheme. A Bayesian \emph{latent} residual  for these variables may be defined as 
 $$\epsilon_{ij} = z_{ij} - \tilde{\underline{\lambda}}_{gj}^T\tilde{\theta}_i.$$
 The sampled values of $z_{ij}$, $ \tilde{\underline{\lambda}}_{gj}$ and $\tilde{\theta}_i$ are used to calculate this residual at each MCMC iteration. If the model fits well such residuals should follow a standard Gaussian distribution. For the nominal SNP variables this residual will be multivariate since two latent dimensions are required to model each nominal SNP. 
 
\begin{figure}[ht]
 \centerline{
 \includegraphics[height = 0.4\textheight, width=0.7\textwidth]{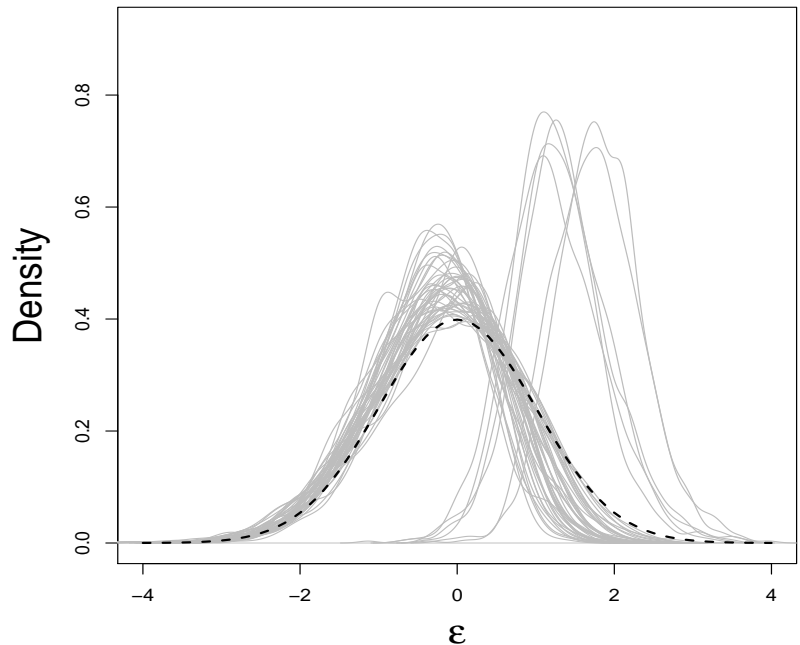}}
 \caption{Density estimates of the Bayesian latent residuals for the {\tt rs17777371} SNP of the \emph{ADD1} gene for 50 randomly selected participants. The standard Gaussian density curve is shown by the black dashed line.}\label{fig:BLRbin}
 \end{figure}

 The Bayesian residuals and latent residuals follow their theoretical distribution reasonably well for the optimal $G=2$, $Q=8$ MFA-MD model. As an example,  Figure~\ref{fig:BLRbin} illustrates kernel density estimates of Bayesian latent residuals corresponding to the \emph{ADD1} ({\tt rs17777371}) SNP for 50 randomly selected participants. The densities are estimated based on the residuals calculated at each MCMC iteration. Curves that do not follow a standard Gaussian distribution correspond to participants whose genotype was unusual given the cluster to which they were assigned. Kernel density estimate plots for other Bayesian residuals and Bayesian latent residuals are provided in the Supplementary Material.
 
 \section{Discussion}
 \label{sec:discussion}
 
The primary focus of the pan European LIPGENE-SU.VI.MAX project is to study the interaction of nutrients and genotype in the metabolic syndrome. Data collected under LIPGENE-SU.VI.MAX are high dimensional and of mixed type, and interest lies in exploring the set of LIPGENE-SU.VI.MAX participants to uncover subgroups with homogeneous phenotypic and genotypic profiles. Examining the link between the resulting clusters and seven-year follow-up MetS diagnosis aids understanding of the role of both phenotypic and genotypic factors in the MetS and provides the opportunity to identify subjects at risk. A clustering method that takes account of different data types and models each one appropriately is therefore necessary. 

While factor analytic methods for data of mixed type and latent factor based clustering methods have already been well developed, the proposed MFA-MD methodology contributes a number of novel advances to the area:
\begin{itemize}
\item the MFA-MD model provides a single, unifying and elegant model for data which notably includes any combination of continuous, binary or nominal response variables. 
\item the MFA-MD approach models nominal response variables in their innate form, rather than requiring a dummy variable representation as is typically necessary in other approaches to clustering nominal response variables. 
\item the variable selection approach permits high dimensional data to be feasibly and efficiently handled, which is theoretically possible but practically challenging for some latent factor models. 
\item the model based approach to clustering and the novel likelihood function approximation facilitates the use of an objective model selection criterion to select the optimal number of clusters and factors rather than relying on subjective heuristic tools.
\end{itemize}


The MFA-MD approach proposed here jointly and elegantly models continuous phenotypic, binary SNP and nominal SNP data, while providing clustering facilities. The suitability of the MFA-MD model for this task is due to its basis in and the relations between a factor analysis model for continuous data, item response theory for binary data and multinomial probit models for nominal data. Further, the parsimonious factor analysis covariance structure is ideal for modelling such high dimensional data. Most of the large number of LIPGENE-SU.VI.MAX data set variables have little to offer in terms of clustering information; a simple and efficient variable selection algorithm is intertwined with the MFA-MD fitting process, thereby highlighting variables that contribute clustering information. This greatly simplifies the task of interpreting the clusters substantively. 

A key aspect of the proposed approach to variable selection is that variables are removed from the model online, thus dramatically reducing the computational burden of fitting the MFA-MD model to high-dimensional data. Several penalisation based variable selection approaches have previously been proposed for latent factor clustering models, for example in \cite{pan2007, galimberti2009, xie2010}; these only consider continuous data in a maximum likelihood framework however. The fact that non-discriminating variables are removed from the MFA-MD model rather than shrinking their associated parameters to zero (meaning all variables are still included in the modelling procedure) ensures the dramatic increase in computational efficiency of the proposed approach. 
 
 As with any clustering problem, of key interest is inferring the number of clusters present in the set of LIPGENE-SU.VI.MAX participants. Standard information criteria approaches in a model based clustering setting involve the evaluation of the observed likelihood function and are not feasible under the MFA-MD model  -- it employs latent variables and evaluation of the observed likelihood function relies on intractable multidimensional integrals. Here an approximation of the observed data likelihood is constructed, and employed in the BIC-MCMC criterion to select both the number of clusters and the dimension of the underlying latent factors in the MFA-MD model. Simulation studies suggest the approximate model selection criterion exhibits desirable performance, as does the variable selection approach taken.

 When applied to the initial mixed phenotypic and genotypic LIPGENE-SU.VI.MAX data, the MFA-MD model uncovers two clusters or `sub-phenotypes' of participants; exploration of the cluster specific parameters suggests one cluster is a `healthy' sub-phenotype and the other an `at risk' sub-phenotype. Both phenotypic and genotypic variables are identified as discriminatory; some are novel discoveries and are indicative of further directions of research. Further, when comparing the resulting clusters to the MetS diagnosis seven years later, the proposed approach out-performs both the use of the standard MetS diagnosis criterion, and the result when clustering using the continuous phenotypic data only, thus emphasising the importance of jointly considering both phenotypic and genotypic profiles when screening for MetS. The proposed MFA-MD approach to clustering provides a global view of the group structure in the set of LIPGENE-SU.VI.MAX participants, but also provides detailed insight to sub-phenotype membership at the participant level, synonymous with the concepts of precision medicine and nutrition. The developed methodology has wide applicability beyond the LIPGENE-SU.VI.MAX study, in any setting seeking to uncover subgroups in a cohort on which high dimensional data of mixed type have been recorded. 
 
 There are many potential areas of future research for the MFA-MD methodology proposed here. Covariate data such as ethnicity and gender are potentially important when studying MetS, and are currently involved in some of the varying MetS diagnosis criteria \cite{alberti05, alberti06}. Incorporating such covariate information in the MFA-MD model could provide understanding of cause-effect relationships in the clustering context. Such information could be incorporated into the MFA-MD model in a mixture of experts framework \cite{jacobs91,gormley08}. 
 
 Within the LIPGENE-SU.VI.MAX cohort a large number of participants were removed from the original data set prior to analysis due to the presence of missing data. To ensure generalisability of the proposed approach it would be advantageous to address such missingness in a more elegant manner. The latent variable and Bayesian origins of the developed model and methodology would allow missing data to be treated as latent variables that can be naturally imputed as part of the MCMC inferential sampling scheme. Such missing data would be required to be missing at random, which was deemed not to be the case in the LIPGENE-SU.VI.MAX cohort. 
 
The approximate model selection criterion developed demonstrated good performance but can be computationally expensive to compute and other approaches have potential merit. Non-parametric approaches to clustering such as the Dirichlet process (or infinite) mixture model \cite{teh10} provide an alternative to the finite mixture approach taken here, and do not require a model selection tool to choose $G$. However, in the case of MFA-MD the value of $Q$ still requires inference; considering an infinite factor model \cite{bhattacharya11} would again avoid the need for a model selection criterion for $Q$, and allow the latent factor dimension to vary across clusters, in a similar manner to that considered in \cite{murphy17}. Such approaches may provide computationally cheaper ways to find the optimal values of $G$ and $Q$ without requiring an expensive grid search.
 
Considering more parsimonious versions of the model \cite{mcnicholas08} would increase modelling flexibility, as would extending the model to include other data types, such as count data, for example. Including such further complexity in the MFA-MD methodology would serve to increase the computational cost of model fitting which, even with the efficiency inducing variable selection procedure, is still somewhat onerous. A variational Bayes approach to estimation of the MFA-MD model \cite{ghahramani99} may have potential in terms of feasibly implementing the model at increased scale and complexity, and may also aid some of the intractable likelihood difficulties. Further, exploring other latent variable representations, for nominal variables in particular, may be fruitful in terms of achieving parsimony and computational efficiency.

\section*{Acknowledgements}

The authors would like to acknowledge the members of the Working Group on Statistical Learning
at University College Dublin and the members of the Working Group on Model-based Clustering at the University of Washington for numerous discussions that contributed enormously to this work. 

The authors would also like to acknowledge the LIPGENE-SU.VI.MAX cohort as the source of these data (FOOD-CT-2003-505944), our cohort co-investigators Serge Hercberg, Denis Larion, Richard Planells, Sandrine Bertrais and Emmanuelle Kesse-Guyot; as well as the LIPGENE-SU.VI.MAX participants.

Finally, the authors would like to thank the reviewers and editors who added considerably to this work through their thorough and thought provoking reviews.

The authors acknowledge LIPGENE-SU.VI.MAX subjects and investigators, funded by European Commission FP6 (FOOD-CT-2003-505944). ICG was supported by Science Foundation Ireland (SFI/09/RFP/MTH2367) and the Insight Research Centre (SFI/12/RC/2289). DMcP was supported by Science Foundation Ireland (SFI/09/RFP/MTH2367).   LB was supported by Science Foundation Ireland (SFI/14/JPI\_HDHL/B3075). HMR was supported by Science Foundation Ireland (SFI/PI/11/1119).

\bibliography{McParlandEtAl.bib}

\end{document}